# Algorithm for Solving Massively Underdefined Systems of Multivariate Quadratic Equations over Finite Fields

Heliang Huang, Wansu Bao*

Zhengzhou Information Science and Technology Institute, Zhengzhou 450000, China

## ABSTRACT

Solving systems of $m$ multivariate quadratic equations in $n$ variables (MQ-problem) over finite fields is NP-hard. The security of many cryptographic systems is based on this problem. Up to now, the best algorithm for solving the underdefined MQ-problem is Hiroyuki Miura et al.'s algorithm, which is a polynomial-time algorithm when $n \geq m(m+3)/2$ and the characteristic of the field is even. In order to get a wider applicable range, we reduce the underdefined MQ-problem to the problem of finding square roots over finite field, and then combine with the guess and determine method. In this way, the applicable range is extended to $n \geq m(m+1)/2$, which is the widest range until now. Theory analysis indicates that the complexity of our algorithm is $O(qn^{\omega}m(\log q)^2)$ when characteristic of the field is even and $O(q2^m n^{\omega}m(\log q)^2)$ when characteristic of the field is odd, where $2 \leq \omega \leq 3$ is the complexity of Gaussian elimination.

### KEYWORDS

underdefined multivariate quadratic equations; multivariate public key cryptosystems; MQ-problem; post quantum cryptographies; algebraic attacks

## 1 INTRODUCTION

The problem of solving multivariate quadratic equations over finite fields is called the MQ-problem. It is well known the general MQ-problem is NP-complete [1]. Multivariate Public Key Cryptosystems (Matsumoto-Imai [2], HFE [3], UOV [4], Flash [5], Rainbow [6] and so on) based on this problem have been expected to secure against the quantum attacks. However, we must note that not all quadratic equations are difficult to be solved. In fact, some of such cryptosystems were already broken and some others of them are weaker than expected when they were proposed. Thus, estimating the hardness of the MQ-problem is important for us to characterize the security of Multivariate Public Key Cryptosystems.

Until now, There are many effective algorithms like Gröbner bases algorithms [7,8,9,10,11] and XL-family [12,13,14] to solve the overdefined ( $m > n$ )

MQ-problem, where *m*, *n* are the numbers of equations and variables respectively. However, not much result seems to be known for finding the solution for the underdefined case ( $n > m$ ). The algorithm for solving underdefined systems of multivariate quadratic equations over finite fields is important for characterizing the security of cryptographic systems. Several multivariate signature schemes are based on the difficulty of solving the underdefined MQ-problem, for example the cryptosystems Flash submitted to Nessie and the Unbalanced Oil and Vinegar scheme (UOV). In fact, when the number of variables *n* is much larger than the number of equations *m*, it is not necessarily difficult to solve these equations. In 1999，Kipnis et al. [4] found a polynomial time algorithm to solve quadratic equations when $n \geq m(m+1)$ and the characteristic of the field is even. Followed by Kipnis et al. , Courtois et al. [15] extended to odd characteristic. Their present version works for roughly about $n \geq 2^{\frac{m}{7}}(m+1)$ and has an exponential complexity in *m*. A polynomial-time algorithm was presented by Hashimoto [16] to solve those over arbitrary fields when $n \geq m^2 - 2m^{3/2} + 2m$ . However, Hiroyuki Miura et al. [17] pointed out that due to some unsolved multivariate equations arisen from the linear transformation, Hashimoto's algorithm doesn't work efficiently. And they presented a polynomial-time algorithm when $n \geq m(m+3)/2$ and the characteristic of the field is even, which is wider than $n \geq m(m+1)$. Thomae et al. [18] modified Kipnis et al.'s algorithm by using Gröbner bases algorithms to close the complexity gap between $n = m$ and $n \geq m(m+1)$, but it is at least a double exponential-time algorithm.

Estimating the difficulty of solving the underdefined MQ-problem, which can provide us a reference for the design of security cryptosystems. For the study of algorithms for solving underdefined MQ-problem, the most central point is to extend the applicable range of the algorithm. In this paper, we face with this difficult problem again, and move forward a step further. Firstly, we reduce the underdefined MQ-problem to the problem of finding square roots over finite field, and then combine with the guess and determine method. In this way, the applicable range is extended to $n \geq m(m+1)/2$, which is the widest range until now. Theory analysis indicates that the complexity of our algorithm is $O(qn^{\omega}m(\log q)^2)$ when characteristic of the field is even and $O(q2^m n^{\omega}m(\log q)^2)$ when characteristic of the field is odd, where $2 \leq \omega \leq 3$ is the complexity of Gaussian elimination. While the complexity of our algorithm is exponential in *m* over the fields of even characteristic, it is still essentially far less than the exhaustive search, especially for large fields.

The paper is structured as follows. In Sect.2 the MQ-problem is described. Our algorithm for solving MQ-problem in the underdefined case is presented in Sect.3. In Sect.4 we compare our proposed algorithm to other known algorithms. Sect.5 concludes this paper.

## 2 THE MQ-PROBLEM

Let $q$ be a power of prime, $F$ be a finite field of order $q$. A MQ-problem over $F$ is given by $m$ equations $f_k = 0$ for polynomial functions $f_k : F^n \to F$ for $1 \le k \le m$ and $a_{k,i,j}, b_{k,i}, c_k \in F$ according to

$$f_k(x_1, \ldots, x_n) = \sum_{1\le i\le j\le n} a_{k,i,j} x_i x_j + \sum_{1\le i\le n} b_{k,i} x_i + c_k \qquad (1)$$

Generally, if we speak of solving such an MQ-problem, we always mean trying to find one solution $(x_1, \ldots, x_n) \in F^n$ such that $f_k(x_1, \ldots, x_n) = 0$ for $1 \le k \le m$. We call $f_k(x_1, \ldots, x_n)$ defined by (1) inhomogeneous. The homogeneous case consists only of terms in $x_i x_j$ and is thus defined by

$$f_k(x_1, \ldots, x_n) = \sum_{1\le i\le j\le n} a_{k,i,j} x_i x_j \qquad (2)$$

Let $\boldsymbol{x} = (x_1, \ldots, x_n)^t \in F^n$, $\tilde{\boldsymbol{x}} = (x_0, x_1, \ldots, x_n)^t \in F^{n+1}$. For any inhomogeneous $f_k(x_1, \ldots, x_n)$, we can construct a homogeneous quadratic $\tilde{f}_k(\tilde{\boldsymbol{x}})$ such that $\tilde{f}_k(1, x_1, \ldots, x_n) = f_k(x_1, \ldots, x_n)$. To ease notation, we restrict to homogeneous system in this article. Obviously, it is must be noted that our algorithm also for inhomogeneous system.

Let *m*, *n* be the numbers of equations and variables respectively, an equation system is called underdefined system when $n > m$, the overdefined system is the opposite.

## 3 PROPOSED ALGORITHM

In the massively underdefined case $n \ge m(m+3)/2$, a polynomial algorithm was developed by Hiroyuki Miura in [17] to solve MQ-problem in case characteristic of the field is even. In this section, we propose a variant algorithm to extend applicable range wider. Similar to Hiroyuki Miura et al.'s algorithm, firstly we reduce the underdefined MQ-problem to the problem of finding square roots over finite field, and then combine with the guess and determine method.

### 3.1 Description of the Proposed Algorithm

Let $(S)$ be the following system:

$$(S)\begin{cases} f_1(x_1,\ldots,x_n)=\sum_{1\le i\le j\le n} a_{1,i,j}x_ix_j=0 \\ \vdots \\ f_m(x_1,\ldots,x_n)=\sum_{1\le i\le j\le n} a_{m,i,j}x_ix_j=0 \end{cases}$$

The main idea of the algorithm consists in using a change of variables such as:

$$\begin{cases} x_1=\alpha_{1,1}y_1+\alpha_{2,1}y_2+\ldots+\alpha_{t,1}y_t+\alpha_{t+1,1}y_{t+1}+\ldots+\alpha_{n,1}y_n \\ \vdots \\ x_n=\alpha_{1,n}y_1+\alpha_{2,n}y_2+\ldots+\alpha_{t,n}y_t+\alpha_{t+1,n}y_{t+1}+\ldots+\alpha_{n,n}y_n \end{cases}$$

whose $\alpha_{i,j}$ coefficients (for $1\le i\le t, 1\le j\le n$) are found step by step, in order that the resulting system $(S')$ (written with respect to these new variables $y_1,\ldots,y_n$) is easy to solve. For $i=1,\ldots,m$, let the polynomial $f_i(x_1,\ldots,x_n)$ be denoted by

$$f_i(x_1,\ldots,x_n)={}^t\boldsymbol{x}F_i\boldsymbol{x}$$

where $F_i, i=1,\ldots,m$ are $n\times n$ matrixes over $F$. An $n\times n$ matrix $T_t$ over $F$ of form (3) is used to transform all the unknowns in "**Step *t*.**" ($t=2,\ldots,m$).

$$T_t=\begin{pmatrix} 1 & 0 & \cdots & 0 & a_{1,t} & 0 & \cdots & \cdots & 0 \\ 0 & 1 & \ddots & \vdots & a_{2,t} & \vdots & & & \vdots \\ \vdots & \ddots & 1 & 0 & \vdots & \vdots & & & \vdots \\ \vdots & & \ddots & 1 & a_{t-1,t} & \vdots & & & \vdots \\ \vdots & & & 0 & 1 & 0 & & & \vdots \\ \vdots & & & \vdots & a_{t+1,t} & 1 & \ddots & & \vdots \\ \vdots & & & \vdots & \vdots & 0 & 1 & \ddots & \vdots \\ \vdots & & & \vdots & \vdots & \vdots & \ddots & 1 & \vdots \\ 0 & \cdots & \cdots & 0 & a_{n,t} & 0 & \cdots & 0 & 1 \end{pmatrix} \qquad (3)$$

where $a_{1,t},\ldots,a_{t-1,t},a_{t+1,t},\ldots,a_{n,t}\in F$.

**Step 1.** Replace $F_i$ by $F_i-c_i^{(1)}F_m$, where $c_i^{(1)}\in F(i=1,\ldots,m-1)$ is a constant such that the (1,1)-element of $F_i-c_i^{(1)}F_m$ is zero. If the (1,1)-element of $F_m$ is zero, exchange $F_m$ for one of $F_1,\ldots,F_{m-1}$ that satisfies the (1,1)-element is not zero.

$$F_i := F_i - c_i^{(1)} F_m \mapsto \begin{pmatrix} \boxed{0} & \\ & * \end{pmatrix}, i \in 1,\ldots,m-1$$

$$f_i := f_i - c_i^{(1)} f_m, i \in 1,\ldots,m-1$$

Then the $F_i, i \in 1,\ldots,m$ has the following form.

$$F_1,\ldots,F_{m-1},F_m \mapsto \underbrace{\begin{pmatrix} \boxed{0} & \\ & * \end{pmatrix},\ldots,\begin{pmatrix} \boxed{0} & \\ & * \end{pmatrix}}_{m-1},\begin{pmatrix} \boxed{*} & \\ & * \end{pmatrix}$$

**Step 2.**(i) Put $f_i(\boldsymbol{x}) := f_i(T_2\boldsymbol{x}), i = 1,\ldots,m$. Determine the $a_{1,2},a_{2,2},\ldots,a_{n,2}$ values in $T_2$ by solving the linear equations such that the coefficients of $x_1x_2$ in $f_1,\ldots,f_m$ are zero.

$$F_1,\ldots,F_{m-1},F_m : \underbrace{\begin{pmatrix} \boxed{0} & \\ & * \end{pmatrix},\ldots,\begin{pmatrix} \boxed{0} & \\ & * \end{pmatrix}}_{m-1},\begin{pmatrix} \boxed{*} & \\ & * \end{pmatrix}$$

$$\mapsto \underbrace{\begin{pmatrix} \boxed{\begin{matrix} 0 & 0 \\ 0 & * \end{matrix}} & \\ & * \end{pmatrix},\ldots,\begin{pmatrix} \boxed{\begin{matrix} 0 & 0 \\ 0 & * \end{matrix}} & \\ & * \end{pmatrix}}_{m-1},\begin{pmatrix} \boxed{\begin{matrix} * & 0 \\ 0 & * \end{matrix}} & \\ & * \end{pmatrix}$$

才

We must note that the (1,2)-element and (2,1)-element of $F_i$ are not always equal to zero. The picture above means that the sum of (1,2)-element and (2,1)-element of $F_i$ is equal to zero for each $i = 1,\ldots,m$.

(ii) Replace $F_i$ by $F_i - c_i^{(2)} F_{m-1}$, where $c_i^{(2)} \in F (i = 1,\ldots,m-2)$ is a constant such that the (2,2)-element of $F_i - c_i^{(2)} F_{m-1}$ is zero. Then we have $f_i := f_i - c_i^{(2)} f_{m-1}, i \in 1,\ldots,m-2$. If the (2,2)-element of $F_{m-1}$ is zero, exchange $F_{m-1}$ for one of $F_1,\ldots,F_{m-2}$ that satisfies the (2,2)-element is not zero.

$$F_1,\ldots,F_{m-1},F_m \mapsto \underbrace{\begin{pmatrix} \boxed{\begin{matrix} 0 & 0 \\ 0 & 0 \end{matrix}} & \\ & * \end{pmatrix},\ldots,\begin{pmatrix} \boxed{\begin{matrix} 0 & 0 \\ 0 & 0 \end{matrix}} & \\ & * \end{pmatrix}}_{m-2},\begin{pmatrix} \boxed{\begin{matrix} 0 & 0 \\ 0 & * \end{matrix}} & \\ & * \end{pmatrix},\begin{pmatrix} \boxed{\begin{matrix} * & 0 \\ 0 & * \end{matrix}} & \\ & * \end{pmatrix}$$

**Step 3.**(i) Put $f_i(\boldsymbol{x}) := f_i(T_3\boldsymbol{x}), i = 1,\ldots,m$. Determine the $a_{1,3},a_{2,3},\ldots,a_{n,3}$ values in $T_3$ by solving the linear equations such that the coefficients of $x_1x_3$ and $x_2x_3$ in $f_1,\ldots,f_{m-1}$ are zero and the coefficient of $x_1x_3$ in $f_m$ are zero.

$$F_1,\ldots,F_{m-1},F_m: \underbrace{\left(\begin{array}{cc|c} 0 & 0 & \\ 0 & 0 & \\ \hline & & * \end{array}\right),\ldots,\left(\begin{array}{cc|c} 0 & 0 & \\ 0 & 0 & \\ \hline & & * \end{array}\right)}_{m-2},\left(\begin{array}{cc|c} 0 & 0 & \\ 0 & * & \\ \hline & & * \end{array}\right),\left(\begin{array}{cc|c} * & 0 & \\ 0 & * & \\ \hline & & * \end{array}\right)$$

$$\mapsto \underbrace{\left(\begin{array}{ccc|c} 0 & 0 & 0 & \\ 0 & 0 & 0 & \\ 0 & 0 & * & \\ \hline & & & * \end{array}\right),\ldots,\left(\begin{array}{ccc|c} 0 & 0 & 0 & \\ 0 & 0 & 0 & \\ 0 & 0 & * & \\ \hline & & & * \end{array}\right)}_{m-2},\left(\begin{array}{ccc|c} 0 & 0 & 0 & \\ 0 & * & 0 & \\ 0 & 0 & * & \\ \hline & & & * \end{array}\right),\left(\begin{array}{ccc|c} * & 0 & 0 & \\ 0 & * & * & \\ 0 & * & * & \\ \hline & & & * \end{array}\right)$$

( ii ) Replace $F_i$ by $F_i - c_i^{(3)}F_{m-2}$, where $c_i^{(3)} \in F(i=1,\ldots,m-3)$ is a constant such that the (3,3)-element of $F_i - c_i^{(2)}F_{m-2}$ is zero. Then we have $f_i := f_i - c_i^{(3)} f_{m-1}, i \in 1,\ldots,m-3$. If the (3,3)-element of $F_{m-2}$ is zero, exchange $F_{m-2}$ for one of $F_1,\ldots,F_{m-3}$ that satisfies the (3,3)-element is not zero.

$$F_1,\ldots,F_{m-1},F_m \mapsto$$

$$\underbrace{\left(\begin{array}{ccc|c} 0 & 0 & 0 & \\ 0 & 0 & 0 & \\ 0 & 0 & 0 & \\ \hline & & & * \end{array}\right),\ldots,\left(\begin{array}{ccc|c} 0 & 0 & 0 & \\ 0 & 0 & 0 & \\ 0 & 0 & 0 & \\ \hline & & & * \end{array}\right)}_{m-3},\left(\begin{array}{ccc|c} 0 & 0 & 0 & \\ 0 & 0 & 0 & \\ 0 & 0 & * & \\ \hline & & & * \end{array}\right),\left(\begin{array}{ccc|c} 0 & 0 & 0 & \\ 0 & * & 0 & \\ 0 & 0 & * & \\ \hline & & & * \end{array}\right),\left(\begin{array}{ccc|c} * & 0 & 0 & \\ 0 & * & * & \\ 0 & * & * & \\ \hline & & & * \end{array}\right)$$

$$\vdots$$

(Repeat the similar operations to "**Step *m***")

With the computation above, we can obtain $F_1,\ldots,F_{m-1},F_m$ of the following form

$$\left(\begin{array}{cccccc|c} 0 & \cdots & \cdots & \cdots & \cdots & 0 & \\ \vdots & \ddots & & & & \vdots & \\ \vdots & & \ddots & & & \vdots & \\ \vdots & & & \ddots & & \vdots & \\ \vdots & & & & 0 & 0 & \\ 0 & \cdots & \cdots & \cdots & 0 & * & \\ \hline & & & & & & * \end{array}\right),\left(\begin{array}{cccccc|c} 0 & \cdots & \cdots & \cdots & \cdots & 0 & \\ \vdots & \ddots & & & & \vdots & \\ \vdots & & \ddots & & & \vdots & \\ \vdots & & & 0 & & \vdots & \\ \vdots & & & & * & 0 & \\ 0 & \cdots & \cdots & \cdots & 0 & * & \\ \hline & & & & & & * \end{array}\right),\left(\begin{array}{cccccc|c} 0 & \cdots & \cdots & \cdots & \cdots & 0 & \\ \vdots & \ddots & & & & \vdots & \\ \vdots & & 0 & & & \vdots & \\ \vdots & & & * & & 0 & \\ \vdots & & & & * & * & \\ 0 & \cdots & \cdots & 0 & * & * & \\ \hline & & & & & & * \end{array}\right),\ldots\ldots,\left(\begin{array}{cccccccc|c} * & 0 & \cdots & \cdots & \cdots & \cdots & \cdots & 0 & \\ 0 & * & \cdots & \cdots & \cdots & \cdots & \cdots & * & \\ \vdots & \vdots & \ddots & & & & & \vdots & \\ \vdots & \vdots & & \ddots & & & & \vdots & \\ \vdots & \vdots & & & \ddots & & & \vdots & \\ \vdots & \vdots & & & & \ddots & & \vdots & \\ \vdots & \vdots & & & & & \ddots & \vdots & \\ 0 & * & \cdots & \cdots & \cdots & \cdots & \cdots & * & \\ \hline & & & & & & & & * \end{array}\right)$$

By transforming the variables with $T_i, i = 2,\ldots,m$, then we have the following

system

$$\begin{cases} \beta_m x_m^2 + \sum_{1\le i\le m} x_i L_{1,i}(x_{m+1},\ldots,x_n) + Q_{1,2}(x_{m+1},\ldots,x_n) = 0 \\ \beta_{m-1} x_{m-1}^2 + Q_{2,1}(x_m) + \sum_{1\le i\le m} x_i L_{2,i}(x_{m+1},\ldots,x_n) + Q_{2,2}(x_{m+1},\ldots,x_n) = 0 \\ \beta_{m-2} x_{m-2}^2 + Q_{3,1}(x_{m-1},x_m) + \sum_{1\le i\le m} x_i L_{3,i}(x_{m+1},\ldots,x_n) + Q_{3,2}(x_{m+1},\ldots,x_n) = 0 \\ \vdots \\ \beta_1 x_1^2 + Q_{m,1}(x_2,\ldots,x_m) + \sum_{1\le i\le m} x_i L_{m,i}(x_{m+1},\ldots,x_n) + Q_{m,2}(x_{m+1},\ldots,x_n) = 0 \end{cases} \quad (4)$$

where each $L_{i,j}$ is an linear polynomial, each $\beta_1,\ldots,\beta_m \in F$ is the coefficient of $x_i^2$ and each $Q_{i,j}$ is a quadratic polynomial.

**Step *m*+1.** Guess the value of $x_1$, and then solve the linear equations $\{L_{i,j}(x_{m+1},\ldots,x_n)=0 \mid i=1,...,m-1; j=2,...,m-i+1\}$. By substituting the solutions $x_{m+1},\ldots,x_n$ into (4), it remains to solve the following system of *m* equations on *m*-1 unknowns $x_2,\ldots,x_m$:

$$\begin{cases} \beta_m x_m^2 = \lambda_m \\ \beta_{m-1} x_{m-1}^2 + Q_{2,1}(x_m) = \lambda_{m-1} \\ \beta_{m-2} x_{m-2}^2 + Q_{3,1}(x_{m-1},x_m) = \lambda_{m-2} \\ \vdots \\ Q_{m,1}(x_2,\ldots,x_m) + L(x_2,\ldots,x_m) = \lambda_1 \end{cases} \quad (5)$$

where $\lambda_1,\ldots,\lambda_m \in F$.

***Remark.*** The reason why we only guess the value of $x_1$ can be easily got from the proof of Theorem 3.1. Briefly, if we don't guess the value of $x_1$, $m(m+3)/2$ linear equations $\{L_{i,j}(x_{m+1},\ldots,x_n)=0 \mid i=1,...,m; j=2,...,m-i+1\}$ should be solved. However, if the value of $x_1$ is guessed and determined, then only $m(m+1)/2$ linear equations $\{L_{i,j}(x_{m+1},\ldots,x_n)=0 \mid i=1,...,m-1; j=2,...,m-i+1\}$ need be solved. The number of linear equations should be solve is reduced by $m$, resulting in a wider application range $n \ge m(m+1)/2$ of the algorithm. Of course, it can be obtained by guessing more than one variable if we don't guess $x_1$, but the computational complexity is increased.

It is very easy for us to solve the quadratic equations of form (5). We solve the first equation in (5) first and then back substitute the solution $x_m$ into other equation.

By substituting the solution $x_m$ to others, we can solve the second equation and then back substitute the solution $x_{m-1}$ into other equation. Repeat the similar operations to the $(m$-$1)$-$th$ equation. The last equation is used to verify the correctness of the solutions. If the solutions are wrong, we should guess another value of $x_1$ until the correct solutions are obtained. This method is called guess and determine method, which is used widely in the area of cryptanalysis.

### 3.2 Theoretical Analysis

In this section, we analyse the conditions and complexity of the proposed algorithm.

First, we give the required conditions of our algorithm in Theorem 3.1.

**Theorem 3.1** For an underdefined multivariate quadratic system of *m* equations in *n* variables over *F*, our algorithm works if $n \geq m(m+1)/2$.

*Proof.* We see that "**Step *t*.**"( $2 \leq t \leq m$ ) requires to solve $(m-t+1)(t-1)+\sum_{i=1}^{t-1} i$ homogeneous linear equations with $n-1$ variables. Then one needs the condition $n-1 \geq (m-t+1)(t-1)+\sum_{i=1}^{t-1} i$ until **Step*m*+1.** That is

$$n-1 \geq \max\{(m-t+1)(t-1)+\sum_{i=1}^{t-1} i\}, t=2,\ldots,m$$

then

$$n-1 \geq \max\{-\frac{1}{2}\{t-(m+\frac{3}{2})\}^2+\frac{1}{2}m^2+\frac{1}{2}m+\frac{1}{8}\}, t=2,\ldots,m$$

Thus, we require $n \geq m(m+1)/2$.

In "**Step *m*+1.**" , the number of linear equations need to be solved is

$$\sum_{t=1}^{m}(m-t)=\frac{1}{2}m(m-1)$$

and the number of variables is $n-m$. Thus, we require $n-m \geq \frac{1}{2}m(m-1)$, that is $n \geq m(m+1)/2$.

In conclusion, our algorithm works if $n \geq m(m+1)/2$.

The algorithm requires to calculate the square roots. When the characteristic of *F* is odd, the probability of existence of square roots is approximately 1/2. Moreover, our algorithm uses only $n \times n$ matrix operations. Theorem 3.2 gives the complexity of our proposed algorithm.

**Theorem 3.2** For an underdefined multivariate quadratic system of *m* equations in *n* variables over *F*, the complexity of our algorithm is

$$\begin{cases} O(qn^{\omega}m(\log q)^2) & (\text{char}\, F \text{ is even}) \\ O(q2^m n^{\omega} m(\log q)^2) & (\text{char}\, F \text{ is odd}) \end{cases}$$

where $2 \le \omega \le 3$ is the complexity of Gaussian elimination.

*Proof.* In this algorithm, about *m* times $n \times n$ matrix operations are calculate. The complexity of this operation is $O(n^{\omega}(\log q)^2)$. In **Step*m*+1**, The probability to guess the correct value of $x_1$ is $q^{-1}$. Therefore, when the characteristic of *F* is even, the complexity of the algorithm is $O(qn^{\omega}m(\log q)^2)$.When the characteristic of *F* is odd, we can find the solution of $x_i, (i = 1, \ldots, m)$ with probability of $2^{-m}$. So the complexity is $O(q2^m n^{\omega} m(\log q)^2)$ when the characteristic of *F* is odd.

# 4 COMPARING EFFICIENCY OF THE ALGORITHMS

In this section, we compare the proposed algorithm with other known algorithms.

## 4.1 When the characteristic of *F* is even

Table 1 presents the applicable ranges of our proposed algorithm and other known algorithms when the characteristic of *F* is even. It is very clear that our approach has a wider applicable range than the others. The applicable ranges of these algorithms are drawn in Fig. 1.

**Table 1.** Applicable ranges of various algorithms when the characteristic of *F* is even

| | **Applicable range** | **Complexity** |
|---|---|---|
| **Our algorithm** | $n \ge m(m+1)/2$ | (poly.) |
| **Kipnis et al.[4]** | $n \ge m(m+1)$ | (poly.) |
| **Courtois et al.[12]** | $n \ge m(m+1)$ | (poly.) |
| **Hiroyuki Miura et al.[17]** | $n \ge m(m+3)/2$ | (poly.) |

When $m = 50$, the applicable range of Kipnis et al.'s algorithm and Courtois et al.'s algorithm is $n \ge 2550$. Hiroyuki Miura et al.'s algorithm requires $n \ge 1325$. Our algorithm further reduce the number of unknowns, and only requires $n \ge 1275$.

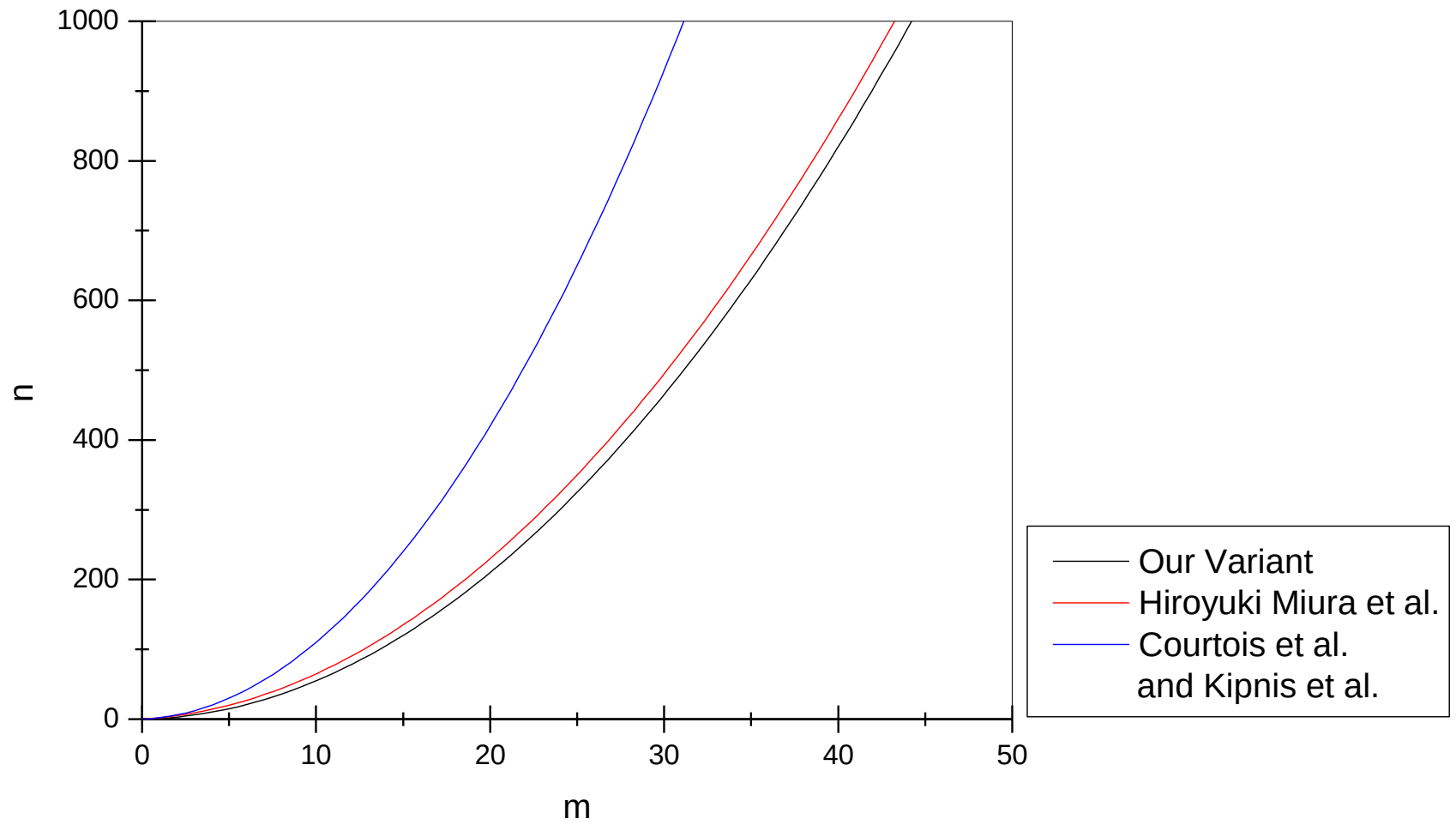

**Fig 1.** Applicable ranges of various algorithms when the characteristic of *F* is even

## 4.2 When the characteristic of *F* is odd

Table 2 presents the applicable ranges of our proposed algorithm and other known algorithms when the characteristic of *F* is odd. The applicable ranges of these algorithms are drawn in Fig. 2.

**Table 2.** Applicable ranges of various algorithms when the characteristic of *F* is odd

| | Applicable range | Complexity |
|---|---|---|
| **Our algorithm** | $n \geq m(m+1)/2$ | (exp.) |
| **Kipnis et al.** | $n \geq m(m+1)$ | (exp.) |
| **Courtois et al.** | $n \geq 2^{\frac{m}{7}}(m+1)$ | (exp.) |
| **Hiroyuki Miura et al.** | $n \geq m(m+3)/2$ | (exp.) |

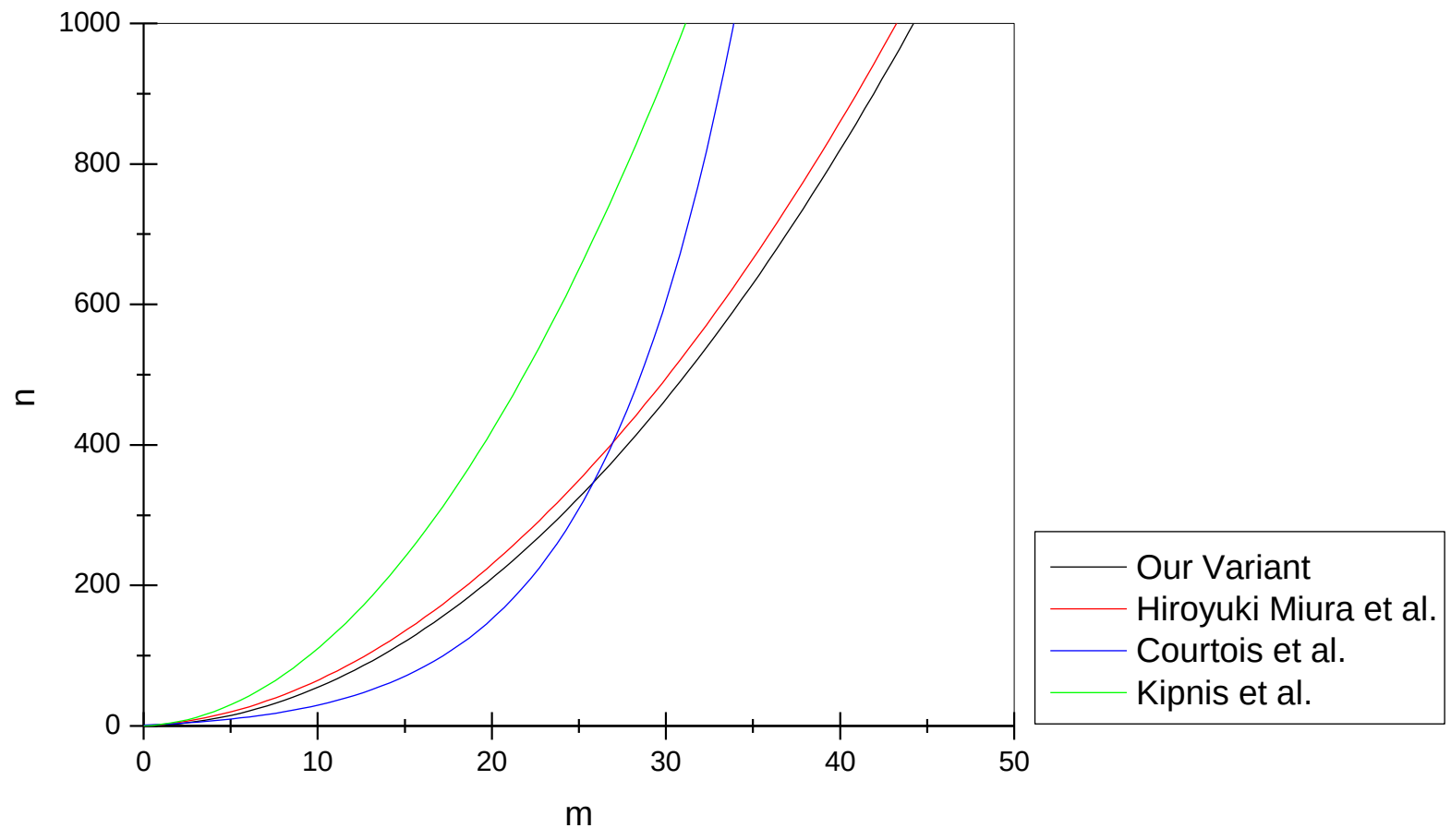

**Fig 2.** Applicable ranges of various algorithms when the characteristic of *F* is odd

From Fig.2, we can see that when $m \leq 25$, Courtois et al.'s algorithm has a wider applicable range than our algorithm. When $m \geq 26$, the applicable range of our algorithm is wider than Courtois et al.'s algorithm. Generally, the number of equations *m* obtained from a multivariate public key cryptosystem is always satisfying $m \geq 26$.

## 5 CONCLUSIONS AND FUTRUE WORK

In this paper, an algorithm for solving the MQ-problem when $n \geq m(m+1)/2$ is presented, where *m*, *n* are the numbers of equations and variables respectively. The algorithm has a widest application range when the characteristic of *F* is even. When the characteristic of *F* is odd and $m \geq 26$, the applicable range of our algorithm is also wider than other algorithms. In fact, the number of equations *m* obtained from a multivariate public key cryptosystem is always satisfying $m \geq 26$. The result can provide a technical support for assessing the security of some cryptosystems. Next, we will continue close the gap between $n = m$ and $n \geq m(m+1)/2$ to make the applicable range wider.

## REFERENCES

1. Garey, M.R., Johnson, D.S.: Computers and Intractability: A Guide to the Theory of NP-Completeness. W.H.Freeman (1979)
2. Matsumoto, T., Imai, H.: Public Quadratic Polynomial-Tuples for Efficient Signature-Verification and Message-Encryption. In: Günther, C.G. (ed.) EURO-CRYPT 1988. LNCS, vol. 330, pp. 419–453. Springer, Heidelberg (1988)
3. Patarin, J.: Hidden Field Equations (HFE) and Isomorphisms of Polynomials (IP):Two New Families of Asymmetric Algorithms. In: Maurer, U.M. (ed.) EURO-CRYPT 1996. LNCS, vol. 1070, pp. 33–48. Springer, Heidelberg (1996)
4. Kipnis, A., Patarin, J., Goubin, L.: Unbalanced Oil and Vinegar Signature Schemes. In: Stern, J. (ed.) EUROCRYPT 1999. LNCS, vol. 1592, pp. 206–222. Springer, Heidelberg (1999)
5. J. Patarin, N. Courtois, L. Goubin, FLASH, a Fast Multivariate Signature Algorithm , in Progress in Cryptology–CT-RSA 2001, D. Nacchache, ed., vol 2020,Springer Lecture Notes in Computer Science, pp. 298-307.
6. Ding, J., Schmidt, D.: Rainbow, a New Multivariate Polynomial Signature Scheme. In: Ioannidis, J., Keromytis, A.D., Yung, M. (eds.) ACNS 2005. LNCS,

vol. 3531, pp. 164–175. Springer, Heidelberg (2005)

7. J. C. Faugère. A new efficient algorithm for computing Gröbner basis (F4). *J. Pure Appl. Algebra*, 1999; 139 (1–3): 61–88.
8. J. C. Faugère. A new efficient algorithm for computing Gröbner bases without reduction to zero (F5), In *ISSAC 2002*, ACM Press: New York, 2002; 75-83.
9. C. Eder and J. Perry. F5C: A variant of Faugère's F5 algorithm with reduced Gröbner bases. *Journal of Symbolic Computation*2010, 45(12):1442–1458.
10. S. H. Gao, Y. Guan and F. Volny IV. A new incremental algorithm for computing Gröbner bases. In *ISSAC 2010*, ACM Press: Munich Germany, 2010; 13-19.
11. Heliang Huang, Wansu Bao. Middle-Solving F4 to Compute Gröbner bases for Cryptanalysis over GF (2). arXiv preprint arXiv:1310.2332 (2013).
12. N. T. Courtois, A.Klimov, J.Patarin, and A. Shamir. Efficient algorithms for solving overdefined systems of multivariate polynomial equations. In *Advance in Cryptology-EUROCRYPT 2000*, pages 392-407, Bruges, Belgium, 2000. Springer.
13. W. S. A. Mohamed, J. Ding, T. Kleinjung, S. Bulygin, and J. Buchmann. PWXL: a parallel Wiedemann-XL algorithm for solving polynomial equations over GF(2). In *Proceedings of the 2nd International Conference on Symbolic Computation and Cryptography (SCC2010)*, pages 89-100, Jun 2010.
14. Heliang Huang, Wansu Bao, Shukai Liu. Parallel Gaussian Elimination for XL-Family over GF(2). *Security and Communication Networks*, 2013.
15. Courtois, N.T., Goubin, L., Meier, W., Tacier, J.-D.: Solving Underdefined Systems of Multivariate Quadratic Equations. In: Naccache, D., Paillier, P. (eds.) PKC 2002.LNCS, vol. 2274, pp. 211–227. Springer, Heidelberg (2002)
16. Hashimoto, Y.: Algorithms to Solve Massively Under-Defined Systems of Multi-variate Quadratic Equations. IEIC E Trans. Fundamentals E94-A(6), 1257–1262(2011)
17. Miura, Hiroyuki, Yasufumi Hashimoto, and Tsuyoshi Takagi. "Extended Algorithm for Solving Underdefined Multivariate Quadratic Equations." Post-Quantum Cryptography. Springer Berlin Heidelberg, 2013. 118-135.
18. Thomae, E., Wolf, C.: Solving Underdetermined Systems of Multivariate Quadratic Equations Revisited. In: Fischlin, M., Buchmann, J., Manulis, M. (eds.) PKC 2012.LNCS, vol. 7293, pp. 156–171. Springer, Heidelberg (2012)